\documentclass[a4paper,11pt]{article}

\usepackage{jinstpub} 
\usepackage{lineno}

\usepackage{graphicx}
\graphicspath{{./}{figure/}}
\usepackage{caption}
\usepackage{subcaption}
\usepackage{enumitem}
\usepackage[dvipsnames]{xcolor}
\usepackage{makecell}
\usepackage{adjustbox}
\usepackage{threeparttable}
\usepackage{multirow}
\usepackage{float}
\usepackage{dblfloatfix}
\usepackage{hyperref}

\title{\boldmath The future of X-ray polarimetry towards the 3-Dimensional photoelectron track reconstruction}

\proceeding{24$^{\text{th}}$ International Workshop on Radiation Imaging Detectors\\
  25-29 June 2022\\
   Oslo, Norway}
   
\author[a,b,c]{Dawoon E. Kim}
\author[a]{Alessandro Di Marco}
\author[a]{Paolo Soffitta}
\author[a]{Enrico Costa}
\author[a]{Sergio Fabiani}
\author[a]{Fabio Muleri}
\author[a]{Ajay Ratheesh}
\author[a,b,c]{Fabio La Monaca}
\author[a]{John Rankin}
\author[a]{Ettore Del Monte}
\author[a]{Alda Rubini}
\affiliation[a]{INAF Istituto di Astrofisica e Planetologia Spaziali, Via del Fosso del Cavaliere 100, 00133 Roma, Italy}
\affiliation[b]{Dipartimento di Fisica, Universit\`a degli Studi di Roma ``La Sapienza'', Piazzale Aldo Moro 5, 00185 Roma, Italy}
\affiliation[c]{Dipartimento di Fisica, Universit\`a degli Studi di Roma ``Tor Vergata'', Via della Ricerca Scientifica 1, 00133 Roma, Italy}

\emailAdd{dawoon.kim@inaf.it}

\abstract{The development of the first X-ray polarimeter, based on the photoelectric effect 20 years ago and implemented thanks to advances in gas amplification structures and readout techniques, had a significant impact in opening a new window for X-ray polarimetry. This system measures the X-ray polarization by reconstructing the initial direction of the photoelectron, emitted by the interaction of an incident photon with an atomic electron, in a gas mixture from an ionization track collected on a two-dimensional plane. However, actual X-ray polarimeters, are still requiring relatively long exposure time and cannot coupled with high effective area mirrors or concentrators. In this context, \emph{the high yield polarimetry experiment in X-rays} (Hype-X) project is currently underway, aiming to improve the sensitivity of the next generation X-ray polarimetry detectors taking advantage of the recent advancements in imaging techniques for high-resolution time projection chambers. In particular, we are evaluating the use of TIMEPIX3 to be applied for the read-out of a gas detector, which will allow us to obtain a three-dimensional image of the photoelectron track. To evaluate the improvement achievable by using a 3D track reconstruction, in this paper, we have reproduced a three-dimensional photoelectron track from a `{\sc Geant4}' Monte Carlo simulation and examined the sensitivity of X-ray polarimetry using a new three-dimensional track reconstruction algorithm. We report the improvement of the modulation factor with three-dimensional track reconstruction as $\sim5\%$ (relative) in the 2-8 keV range and $\sim17\%$ (relative) in the 2-4 keV range compared to the current two-dimensional polarimetry system. This is equivalent to add a further telescope to the three-telescope systems now employed in space on board the IXPE mission.
}


\arxivnumber{} 

\begin{document}
\maketitle 
\flushbottom

\vspace{-0.8cm}
\section{Introduction}
\vspace{-0.1cm}
X-ray polarimetry has been considered for a long time as a prominent tool for exploring the hidden properties of light emanating from the energetic and innermost regions of astrophysical sources in the universe. Nevertheless, due to the lack of suitable sensitive instrumentation, polarimetry had remained a veiled field in X-ray Astronomy compared to other wavelengths. In this context, the development of the first photoelectron-based X-ray polarimeter \cite{2001Natur.411..662C} had a big impact to open a new window for X-ray astrophysics. This system measures the X-ray polarization by reconstructing the initial direction of the photoelectron produced by the interaction of an incident photon with an atomic electron in a gas mixture (i.e., DME; Dimethyl ether) from the collected two-dimensional ionization track. The first polarimeter built following this concept, the Gas Pixel Detectors (GPD) \cite{2001Natur.411..662C, 10.1117/12.459380, 2021APh...13302628B}, has been installed in the CubeSat mission polarimeter light (\textit{PolarLight}) \cite{2019ExA....47..225F} and is utilized as the focal plane of the Imaging X-ray Polarimetry Explorer (\textit{IXPE}) \cite{weisskopf22} that significantly expanding this field \cite[see e.g.,][]{2022Natur.611..677L}. 

Nonetheless, one of the major limitation of X-ray polarimeters is the dead time, which restricts the effective area of the mirrors or concentrators to be coupled with the detector. Therefore, in order to expand our perspective and delve deeper into the universe with X-ray polarimetry, enhancing the performance of the GPD is essential, as it will advance the sensitivity of X-ray polarimetry. In this context, \emph{the high yield polarimetry experiment in X-rays} (Hype-X) project aims to improve the sensitivity of the X-ray polarimeter by combining the capabilities of the GPD thanks to recent advancements in detectors with imaging and timing properties, and ultimately developing the next generation of X-ray polarimetry detectors.

The application of the TIMEPIX3 \cite{2014JInst...9C5013P} as the ASIC of a gas detector, which can provide single electron sensitivity and 3D tracking, will allow to perform a time-resolved X-ray polarimetry with virtually no dead time. The TIMEPIX3 can measure simultaneously \emph{time of arrival} (TOA) and \emph{time-over-threshold} (TOT) for each pixel, effectively providing charge and time measurement for three-dimensional tracking reconstruction. This capability will not only improve sensitivity by creating a three-dimensional image of the photoelectron track but also they will allow it to sustain a high counting rate, enabling the use of large effective areas for collecting a large number of photons thanks to its parallel read-out system. In this regard, it is crucial to develop a new three-dimensional track reconstruction algorithm for determining the initial direction of the photoelectron based on the three-dimensional track image. Therefore, in this paper, we aim to present our initial attempt at this new algorithm for three-dimensional track reconstruction. The reduction of the dead time can also be very effective on the sensitivity but this is case dependent and is not the object of the present study.
	
In this work, we present a three-dimensional track reconstruction algorithm and its application to `{\sc Geant4}' Monte Carlo simulation \cite{AGOSTINELLI2003250} to compare the sensitivity for X-ray polarimetry transitioning from 2D to 3D track reconstruction. In \S\ref{sec:basic_xpol}, we provide a brief introduction to the basic concept of the current two-dimensional photoelectric X-ray polarimeter. Next, we introduce our three-dimensional charge distribution reconstruction process and initial attempt to determine polarization properties by establishing a new three-dimensional moments track reconstruction algorithms in \S\ref{sec:reproduction} and \S\ref{sec:reconstruction}. In \S\ref{sec:results}, we present a comparison of X-ray polarization properties between two different methods, aiming to quantify the prospective polarization improvement with the new method by simulating a number of photoelectron track events in different energy bands.

\vspace{-0.1cm}
\section{Basic concept of photoelectric X-ray polarimeter}\label{sec:basic_xpol}
\vspace{-0.1cm}

   \begin{figure}[t]
   \centering
   \includegraphics[width=0.28\hsize]{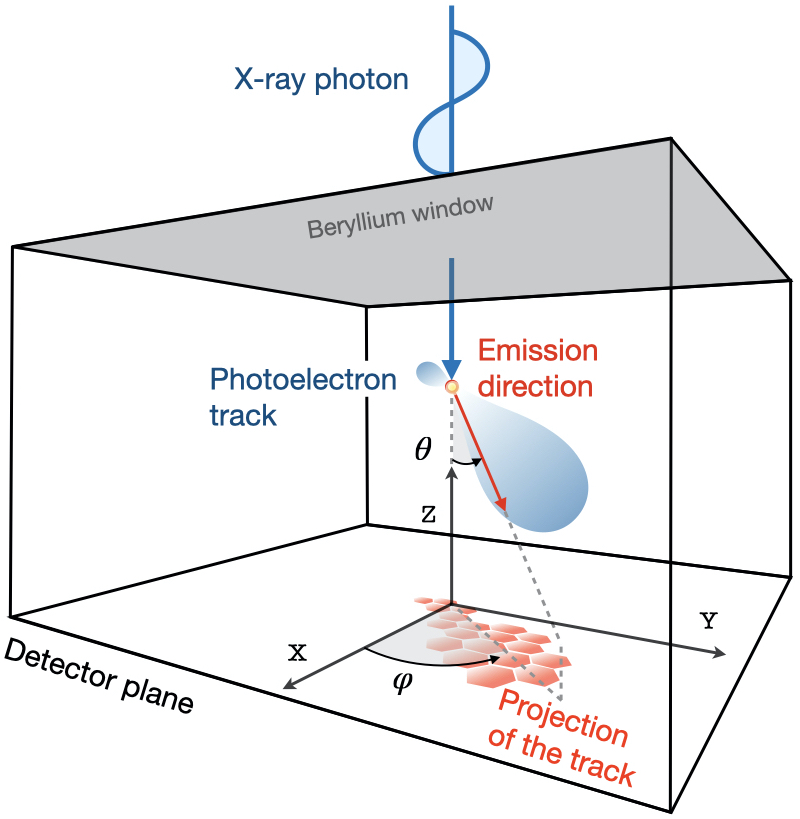}
   \quad
   \includegraphics[width=0.55\hsize]{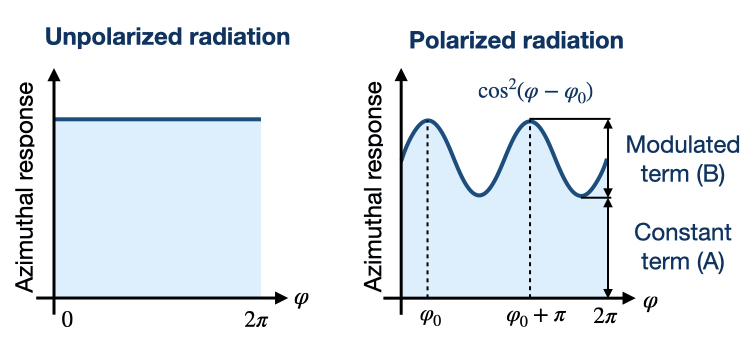}
      \caption{Left: schematic view of a photoelectric X-ray polarimeter. The incoming X-ray photon is absorbed in the gas mixture inside the detector and creates a photoelectron track. The initial direction of this track, which is expressed by $\theta$ and $\varphi$, is correlated with the electric field direction of the absorbed photon. Right: histogram of emission angles of the photoelectrons (\emph{modulation curve}). In the case of unpolarized sources, the distribution is flat, while polarized sources produce cosine square modulation.}
         \label{fig:photoelectric_polarimeter}
   \end{figure}
   
The Gas Pixel Detector analyzes the polarization information based on the initial direction of the photoelectron produced by the interaction between the X-ray photon and the gas mixture inside the detector. Figure \ref{fig:photoelectric_polarimeter} illustrates the concept of a photoelectric polarimeter. In detail, when the X-ray photon interacts with absorption material in the detector, a photoelectron is ejected from the interacting atom. The angular distribution of the ejection direction follows the photoelectric differential cross-section:
\begin{equation}\label{eq:cross}
\small
{{\rm{d} \sigma_{Ph}}\over{\rm{d} \Omega}} \propto \cos^2 \varphi{{{\sin} ^2 \theta}\over{\left( 1 - \beta {\cos \theta}\right)^4}}
\end{equation}
\noindent Where $\varphi$ and $\theta$ represent the azimuthal and polar angles of the photoelectron, respectively, and $\beta$ is the photoelectron velocity in units of the speed of light. The emission direction of the photoelectron is most probably aligned along the direction of the incoming beam electric field. The modest bending forward given by the term at the denominator at our energies is almost negligible. In the subsequent step, the photoelectron interacts in the gas mixture and induces an ionization track inside the detector system until it loses all the energy and stops this process. Then, the drift field from the window to the multiplication stage close to the ASIC readout plane forces the track to move to the Gas Electron Multiplier and is eventually collected by the ASIC CMOS chip, resulting in a two-dimensional projected photoelectron track. However, the photoelectron interacting in the gas mixture loses information about its initial emission during the subsequent processes in the detector. Thereby, the current GPD system employs a reconstruction algorithm based on the moment analysis proposed by \cite{10.1117/12.459380,  2022AJ....163..170D} to measure the emission direction of the photoelectron and the absorption position. 
   
In order to examine the source polarization properties, it requires collecting enough photoelectrons to build the so-called \emph{modulation curve}, that is, the histogram of the photoelectron directions of emission (see left panel of Figure \ref{fig:photoelectric_polarimeter}). The modulation curve shows different shapes according to the presence or absence of polarization of the incoming photons. For instance, a polarized photon beam presents an anisotropic distribution according to the photoelectron emission direction, and an unpolarized one shows a uniform distribution. Hence, by the amplitude and phase of the modulation shown in the curve, we can evaluate the source polarization properties. We first fit the modulation curve with the function:
\begin{equation}\label{eq:modulation}
\small
	{\mathcal {M}} \left( \varphi \right) = {A} + B{\cos^2} \left(\varphi -\varphi_0 \right),
\end{equation}
\noindent where $A$ express constant term and $B$ is the amplitude of the modulation curve. Then, we calculate the polarization degree of the X-ray source (PD) according to: 
\begin{equation}\label{eq:modul_factor}
\small
{\rm PD} = {{1}\over{\mu}}{{B}\over{2A+B}},
\end{equation}
where $\mu$ represents the modulation amplitude measured by the detector for 100 \% polarization photon beam. In this regard, $\mu$ can be considered as a scale factor to evaluate the polarization of the source. The polarization angle (PA) is derived as the phase of a $\cos^2$ fit of the modulation curve. 

\vspace{-0.1cm}
\section{Reconstruction of 3-dimensional charge distribution from the projected two-dimensional track image}\label{sec:reproduction}
\vspace{-0.1cm}

\begin{figure}[t!]
\centering         
\includegraphics[width=0.605\columnwidth]{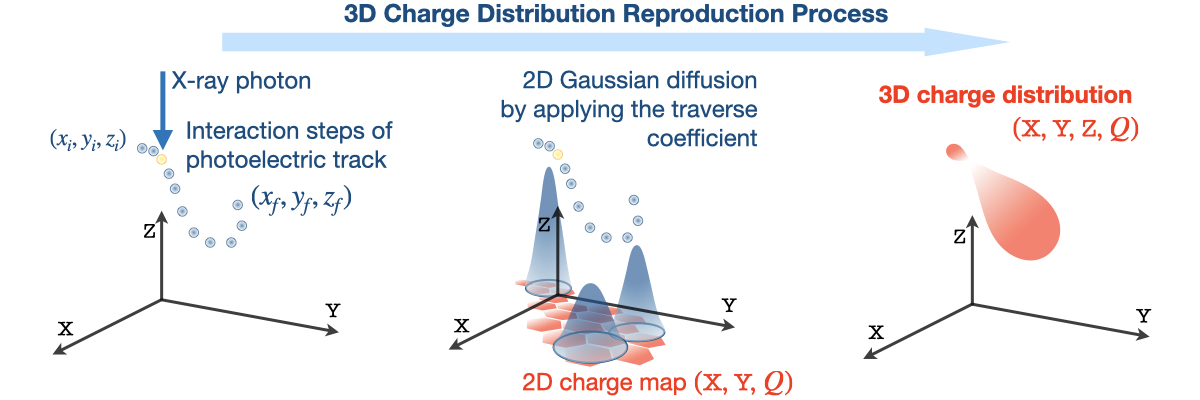}
\caption{Concept of three-dimensional charge distribution reproduction. The yellow and blue dots indicate the original interaction points of the photoelectron track. The blue conical shapes illustrate the Gaussian diffusion process and the red hexagons depict the activated charged pixel on the ASIC in two dimension space. The rightmost red track describes the final three-dimensional charge distribution.
}\label{fig:3DPE_process}
\end{figure}

The outcomes of the next generation polarimeter will be a charge distribution of the photoelectron track in three-dimensional space ($\texttt{X}$, $\texttt{Y}$, $\texttt{Z}$, $Q$). In this work, we created the three-dimensional charge distribution by extending the `{\sc Geant4}' simulation \cite{AGOSTINELLI2003250}. The {\sc Geant4} simulation provides the original interaction points coordinate of the track $(x_i, y_i, z_i)$ and the two-dimensional charge map ($\texttt{X}$, $\texttt{Y}$, $Q$ in Figure \ref{fig:3DPE_process}) following a Gaussian diffusion process with a transverse diffusion coefficient ($\sigma_t=68 \,{\rm\mu m\,cm^{-1/2}}$; calculated with \href{https://garfield.web.cern.ch/garfield/}{Garfield} based on pure DME at 800 mbar) expressed as:
\begin{equation} \label{eq:gaussian}
\small
f (x, y) = \frac{1}{\sqrt{2 \pi \sigma^2}} \exp\left[-\frac{(x-\mu_x)^2 + (y-\mu_y)^2}{2 \sigma^2}\right],
\end{equation}
\noindent where $\sigma$ is $\sigma_t \sqrt{z_i}$. Hence, we reproduce the three-dimensional charge distribution by tracing the contributions of each interaction point on the two-dimensional charge map and assigning the averaged $z_i$ to each pixel. Figure \ref{fig:3DPE_process} illustrates the concept of this reproduction process. In this process, we also calculated the standard deviation of $Z$ direction on each charged pixel ($\sigma_i$). This quantity is exploited in the track reconstruction process described in \S\ref{sec:reconstruction}. 

\vspace{-0.1cm}
\section{3-dimensional moments track reconstruction algorithm}\label{sec:reconstruction}
\vspace{-0.1cm}

In order to determine the initial direction of a three-dimensional track, we developed a new reconstruction algorithm, based on a two-step analysis procedure, using a regression method and three-dimensional rotation. The basic concept of this algorithm is to apply the standard two-dimensional moment analysis method which was developed and applied by the \textit{IXPE} collaboration \cite{10.1117/12.459380, 2022AJ....163..170D}. 

\begin{figure}[t]
\centering
\includegraphics[width=.85\textwidth]{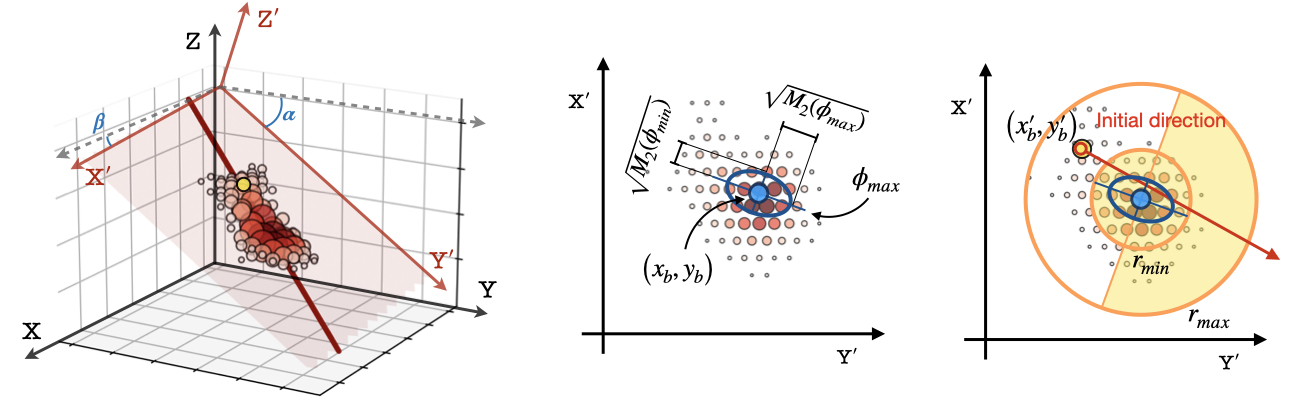}
\caption{The three-dimensional photoelectron track reconstruction process. Left: the three-dimensional charge distribution with the determined regression plane. The yellow dot indicates the absorption point, and the red line represents the principal axis of the initial direction. Right: a detailed view of the two-dimensional moment analysis.\label{fig:moment}} 
\end{figure}

As a first step, we determine the regression plane of the track ($\texttt{Z}=A\texttt{X}+B\texttt{Y}+C$; shown as a red plane in the left panel of Figure \ref{fig:moment}), which represents the averaged three-dimensional positions of the track, by the least chi-square method according to: 
\begin{equation}\label{eq:regression}
\small
\chi^2 = \sum_{i=1}^n {\left( { \tt{Z} - A \tt{X} - B \tt{Y} -C \over \sigma_i} \right)^2,  \qquad  \frac{\partial \chi^2}{\partial A}= \frac{\partial \chi^2}{\partial B} = \frac{\partial \chi^2}{\partial C}= 0,
}
\end{equation}
\noindent where $\texttt{X}$, $\texttt{Y}$, and $\texttt{Z}$ are three dimension coordinates of each charged pixel. $\sigma_i$ indicates the standard deviation into the $\texttt{Z}$ direction of each charged pixel (described in \S\ref{sec:reproduction}).

When the regression plane is determined, we projected the track onto it by replacing $\texttt{Z}$ of each charged pixel with $\texttt{Z}'$. Then, we rotate the $\texttt{X}\texttt{Y}$ plane for applying the two-dimensional moment analysis. The rotation was conducted by constraining the rotation matrices expressed:
\begin{equation}\label{eq:3d_trans1}
\small
R_x = \begin{bmatrix}
    1  & 0 & 0  \\
    0  & \cos{\alpha} &  -\sin{\alpha} \\
    0  & \sin{\alpha} &  \cos{\alpha} \\
    \end{bmatrix}, \quad
R_y = \begin{bmatrix}
    \cos{\beta}  & 0 & -\sin{\beta} \\
    0  & 1 &  0 \\
    \sin{\beta} & 0 &  \cos{\beta} \\
    \end{bmatrix},\quad
R_z = \begin{bmatrix}
    \cos{\gamma}  & \sin{\gamma} & 0 \\
    -\sin{\gamma}  & \cos{\gamma} &  0 \\
    0  & 0 &  1  \\
   \end{bmatrix},
\end{equation}
\noindent where each rotation matrix ($R_x$, $R_y$, $R_z$) represents $\alpha$, $\beta$, and $\gamma$ angle rotation around the corresponding axes $x$, $y$, and $z$. These rotation matrices allow to convert the track properties into the rotated from the (T$'$) and original frame (T) and vice-versa:
\begin{equation}\label{eq:3d_trans2}
\small
   {\bf R} = R_{x}\cdot R_{y}\cdot R_{z}, \qquad T' = {\bf R} \cdot T \leftrightarrow {\bf R}^{-1}T' = T
\end{equation}

The angle for each rotation matrix was calculated with respect to each $\texttt{X}$ and $\texttt{Y}$ axis ($\alpha$ and $\beta$ in Figure \ref{fig:moment}), however, here in the case of $R_{z}$ was constrained as an identity matrix because the rotation is performed to the two-dimensional plane.
Next, based on the two-dimensional moment analysis method \cite{10.1117/12.459380, 2022AJ....163..170D}, we find the principal axis of the initial direction on the $\texttt{X}'\texttt{Y}'$ plane: 
\begin{enumerate}[noitemsep]
	\item Finding a barycenter $(x_b, y_b)$ of charge distribution based on the measured $\texttt{X}'$ and $\texttt{Y}'$ position and each charge density $Q$.
	\item Identifying the first principal axis $(\phi_{max})$ of the charge distribution through a second moment analysis centered on the barycentric position.
	\item By analyzing the third moment, to distinguish the head and tail area of the track from the Bragg peak at the end of the photoelectron path, where a large amount of energy is collected (yellow region in Figure \ref{fig:moment}).
	\item Only considering the track starting area (white half-ring area opposite to the yellow region in Figure \ref{fig:moment}), repeat steps 1 and 2 to estimate the absorbed point $(x'_b, y'_b)$ of the incident X-ray photon (yellow dot with red outline in right panel of Figure \ref{fig:moment}) and find the final principal axis (red arrow line in right panel of Figure \ref{fig:moment}) of the initial direction based on the absorbed point in three dimension space.
\end{enumerate}
\noindent In step 3, the initial portion of the track is empirically determined as a region that is $1.5 -3.0$ times the length of the major axis of the ellipse ({\footnotesize $\sqrt{M_2(\phi_{max})}$}), so that we remove the Bragg peak region. In addition, the absorption points $(x'_b, y'_b)$ measured by the reconstruction process allow to reproduce the image of the observed source. At this stage, the total charge density of the single track is converted into the energy of the absorbed photon.

Lastly, we rotate the principal axis of the initial direction (red arrow line in the right panel on Figure \ref{fig:moment}) back into the original frame (red solid line in the left panel on Figure \ref{fig:moment}) and measure the $\theta$ and $\varphi$ of the photoelectron in the three-dimensional space.

\vspace{-0.1cm}
\section{Sensitivity enhancement by the new 3-dimensional moments track reconstruction}\label{sec:results}
\vspace{-0.1cm}

\begin{figure}[t]
\centering
\includegraphics[width=.31\textwidth]{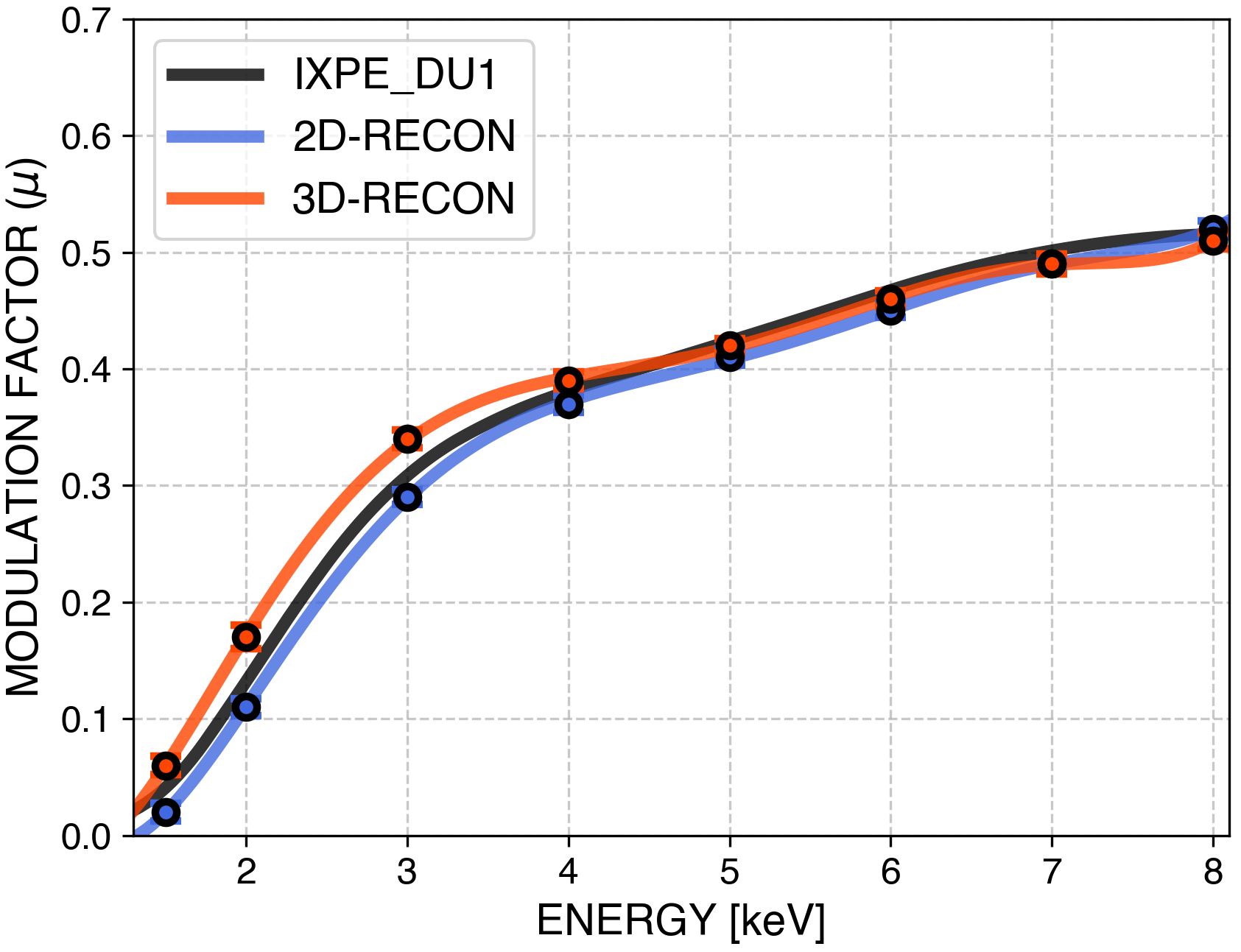}
\quad
\includegraphics[width=.31\textwidth]{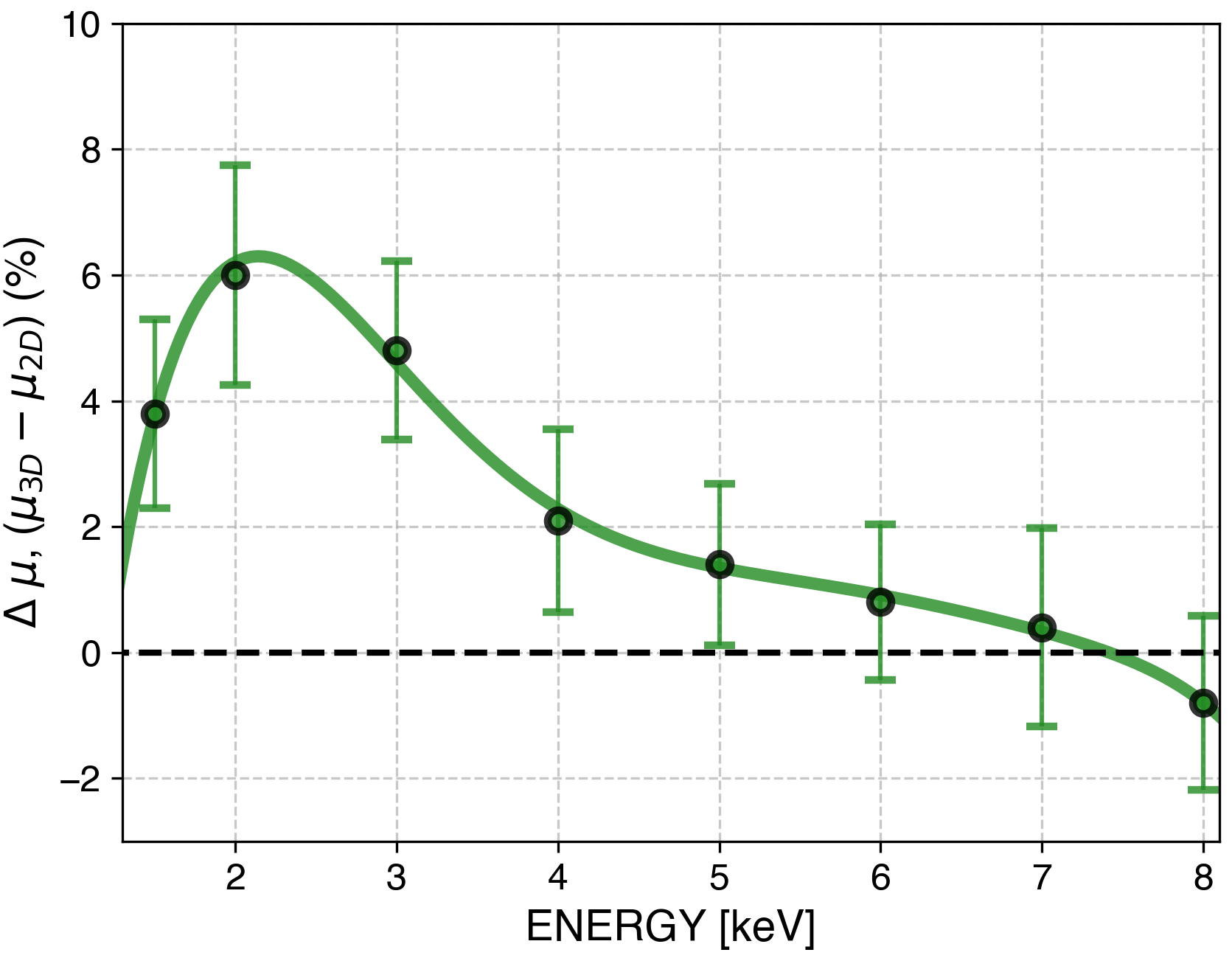}
\quad
\includegraphics[width=.31\textwidth]{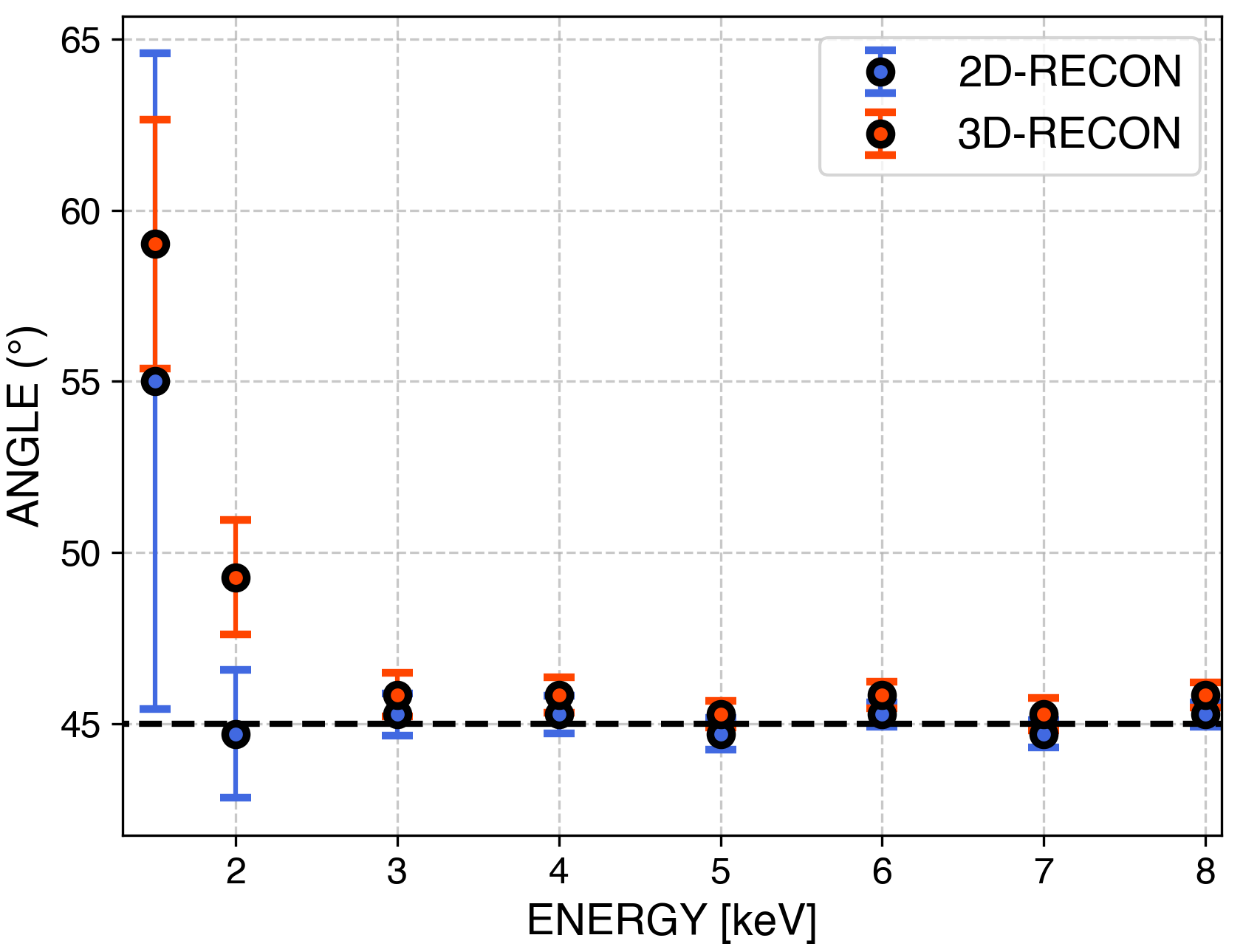}
\caption{Comparison of modulation factors (left, middle) and polarization angles (right) between the two-dimensional (blue) and three-dimensional (red) reconstruction processes as functions of energy. In the left panel, the black solid line represents the measured modulation factor for the photoelectric polarimeters on-board \textit{IXPE} \cite{2021AJ....162..208S, 2022AJ....163..170D, 2022AJ....164..103D}, while blue and red lines are obtained by simulating either a 2D or a 3D photoelectron reconstruction. Middle panel shows the modulation factor difference ($\Delta \mu$) between $\mu_{3D}$ and $\mu_{2D}$ in percentage unit. In the right panel, the dashed line denotes the simulation input value of the PA, which is 45 degree.}\label{fig:results}
\end{figure}

In order to quantify the improvement in the sensitivity to X-ray polarization that can be obtained thanks to the 3D track reconstruction, we simulated 50000 events using the {\sc Geant4} simulation. For the simulation input, we considered monochromatic sources, having the energies reported in Table 1, in a 1.5-8 keV setting, with a PD of 100\% and a PA of 45 degrees. We estimated the modulation factor and determined the PA based on modulation curves for each case. The obtained values are summarized in Table \ref{tab:results} and Figure \ref{fig:results} shows a comparison of these values as a function of energy. 

\begin{table}[t!]
\footnotesize
\caption{The modulation factor and polarization angle are estimated at different energy levels in two-dimensional (2D) and three-dimensional (3D) reconstruction. \label{tab:results}}\centering
\begin{tabular}{ccc}
\hline\hline           
\noalign{\smallskip}
Energy ($\rm{keV}$) & Modulation factor ($\mu$) & PA ($^\circ$)\\
 & 2D \qquad\qquad\qquad\quad 3D & 2D \qquad\qquad\qquad\quad 3D\\
\hline
1.5 & 0.022 $\pm$ 0.007 \qquad 0.060 $\pm$ 0.008& 55.00 $\pm$ 9.59  \qquad 59.01 $\pm$ 3.63 \\
2.0 & 0.112 $\pm$ 0.007 \qquad 0.172 $\pm$ 0.010& 44.69 $\pm$ 1.87  \qquad 49.27 $\pm$ 1.67 \\
3.0 & 0.291 $\pm$ 0.006 \qquad 0.339 $\pm$ 0.008& 45.26 $\pm$ 0.62  \qquad 45.84 $\pm$ 0.64 \\
4.0 & 0.368 $\pm$ 0.007 \qquad 0.389 $\pm$ 0.007& 45.26 $\pm$ 0.55  \qquad 45.84 $\pm$ 0.52 \\
5.0 & 0.411 $\pm$ 0.007 \qquad 0.425 $\pm$ 0.006& 44.69 $\pm$ 0.46  \qquad 45.26 $\pm$ 0.39 \\
6.0 & 0.451 $\pm$ 0.006 \qquad 0.459 $\pm$ 0.007& 45.26 $\pm$ 0.35  \qquad 45.84 $\pm$ 0.39 \\
7.0 & 0.488 $\pm$ 0.007 \qquad 0.492 $\pm$ 0.009& 44.69 $\pm$ 0.40  \qquad 45.26 $\pm$ 0.47 \\
8.0 & 0.518 $\pm$ 0.007 \qquad 0.510 $\pm$ 0.007& 45.26 $\pm$ 0.36  \qquad 45.84 $\pm$ 0.37 \\
\hline
\end{tabular}
\end{table}

As a result, we found that the three-dimensional reconstruction exhibited better performance compared to the current two-dimensional algorithm, with an improved modulation factor of $\sim$5\% (relative) in the 2-8 keV range and $\sim$17\% (relative) in the 2-4 keV range. This improvement, being the quantum efficiency equal in the two methods, corresponds to the overall improvement of the performances. This means, for instance, that we could reduce the exposure time as $\sim$9.0 days in 2-8 keV and $\sim$7.3 days in 2-4 $\rm{keV}$ for 10 days observation achievement with the 2D algorithm. In particular, the $\sim$17\% improvement in the 2-4 keV range is physically equivalent to add an additional telescope (a four-telescope system) to a hypothetical three-telescope system like \textit{IXPE}. However, we also note that the final angle measurement deviates in the soft energy band, mainly outside the IXPE band, when using the current three-dimensional reconstruction. This issue should be addressed in the enhanced reconstruction algorithm in the future.

\vspace{-0.1cm}
\section{Conclusions}\label{sec:conclusions}
\vspace{-0.1cm}
The three-dimensional reconstruction process of the photoelectron track allows for an improvement of the performances, especially at low energies, where the tracks are shorter and appear almost circular when projected onto 2D planes. The three-dimensional reconstruction requires the knowledge of the z coordinate, this can be obtained by timing information provided from the TIMEPIX3 \cite{2014JInst...9C5013P}, thanks to its high time resolution. Currently, a collaboration between INAF/IAPS and the University of Bonn is developing this next-generation detector system based on this technology. Additionally, a new {\sc Geant4} Monte Carlo simulation software is in development at INAF/IAPS to contribute to modeling the response of this new detector.

\acknowledgments
{\footnotesize
\textit{Acknowledgments.} The authors acknowledge funding from the Minister of University and Research PRIN 2020MZ884C ``Hype-X: High Yield Polarimetry Experiment in X-rays'' and the INAF MiniGrant ``New generation of 3D detectors for X-ray polarimetry: simulation of performances''}

\bibliographystyle{JHEP}
{
 \bibliography{ref.bib}
}

\end{document}